\documentstyle[12pt]{article}
\catcode`\@=11
\def \be {\begin{equation}}
\def \ee {\end{equation}}
\def \beq {\begin{equation}}
\def \eeq {\end{equation}}
\def \ba {\begin{eqnarray}}
\def \ea {\end{eqnarray}}
\def \baa {\begin{eqnarray*}}
\def \eaa {\end{eqnarray*}}
\def\bea{\begin{eqnarray}}
\def\eea{\end{eqnarray}}

\def\beq{\begin{equation}}
\def\eeq{\end{equation}}
\def\ba{\beq\new\begin{array}{c}}
\def\ea{\end{array}\eeq}
\def\be{\ba}
\def\ee{\ea}

\begin{document}


\setcounter{footnote}0



\begin{flushright}
{ ITEP/TH-60/05\\
FITP-05/45}
\end{flushright}

\vspace{1.3cm}

\begin{center}
{\Large \bf From Yang-Mills Lagrangian to MHV Diagrams }
\end{center}

\begin{center}

{ { \bf A.~Gorsky\,$^{a,b}$}} and
{ \bf    A.~Rosly$^{a}$}

\vspace{0.3cm}

$^a${\it Institute of Theoretical and Experimental Physics, Moscow
117259, Russia}\\
$^b${\it  William I. Fine Theoretical Physics Institute,
University of Minnesota,
Minneapolis, MN 55455, USA}

\vspace*{.45cm}

{\large\bf Abstract}
\end{center}

We prove the equivalence of a recently suggested MHV-formalism to
the standard Yang-Mills theory. This is achieved by a formally
non-local change of variables. In this note we present the
explicit formulas while the detailed proofs are postponed to a
future publication.

\section{Introduction}

Recently, a new approach to the perturbative calculations in
Yang-Mills (YM) theory has been suggested by Cachazo, Svr\v cek and
Witten (CSW) \cite{csw}. In this new formalism, the vertices are
obtained from the so-called MHV-amplitudes (i.e.\ the amplitudes
maximally violating the helicity) by a suitable continuation off
shell. This technique was shown to reproduce all known gluon tree
amplitudes and predicts a number of new results \cite{xxx}. The
successful generalization for the one-loop amplitudes has been
also developed \cite{one-loop} although a new additional vertex
has to be added at one-loop level in YM theory without
supersymmetry. The MHV-like diagrams for the gravity case have
been formulated as well \cite{gravity}. A complete list of
references can be found in \cite{cs}.

In this paper we address the question of equivalence between the
MHV diagrams and the conventional YM perturbation theory
expansion. The MHV diagram rules can of course be  described
with help of an action functional, which we call the CSW action.
It turns out that there exists a change of variables transforming
the standard YM action to the CSW action. The formula for such a
change of variables is obtained as follows. First, we recall a
certain solution to the self-duality equation which serves for a
swift derivation  of the MHV-amplitudes \cite{bardeen,rs}. This
self-dual gauge field can be continued off shell in the spirit of
ref.\ \cite{csw} and provides  very explicit change of variables
which brings YM Lagrangian in the light-cone gauge into the form
of CSW Lagrangian. At present, we can check this  by a
brute-force calculation only and  feel that a better, more
conceptual understanding of our result is needed.  This is
despite the fact that the formula for the change of variables is
perfectly explicit and the geometrical origin of the self-dual
solution behind it seems to be well understood. Therefore, we
give here those explicit formulas and postpone the detailed proofs
to a future publication.

The paper is organized as follows. First, we remind the MHV
diagram rules (Section 2) and present the YM action in the
light-cone gauge (Section 3). Then, in Section 4 we describe a
solution to the self-duality equation which is relevant to the
MHV-amplitudes. A change of variables in the light-cone YM action,
which renders it to the CSW action, is introduced in Section 5.
Some open questions are mentioned in the concluding Section.

\section{MHV diagrams}

Let us remind main points concerning  MHV diagrams. The basic
ingredient is  MHV $(--,+\dots +)$ amplitude describing the tree
scattering
of two gluons of negative helicity and arbitrary number of positive
helicity gluons. The amplitude turns out to be a simple
rational function of on-shell momenta of massless particles and reads
as \cite{parke,bg}
\beq
A(1^{-},2^{-},3^{+}\dots ,n^{+})= g^{n-2}
\frac{\langle1,2\rangle^4}{\langle1,2\rangle\langle2,3\rangle\dots \langle n,1\rangle}
\eeq
where the on-shell momentum of massless particle in the standard spinor
notations
reads as $p_{a\dot{a}}= \lambda_a\tilde{\lambda}_{\dot{a}}$,
$\lambda_a$ and $ \tilde{\lambda}_{\dot{a}}$
are positive and negative helicity spinors. Inner products
in spinor notations read as
$\langle\lambda_1,\lambda_2\rangle= \epsilon_{ab}\lambda_1^{a}\lambda_2^{b}=\langle1,2\rangle$
and $[\tilde{\lambda}_1,\tilde{\lambda}_2]=
\epsilon_{\dot{a}\dot{b}}\tilde{\lambda}_1^{\dot{a}}\tilde{\lambda}_2^{\dot{b}}$.
These amplitudes were interpreted as correlators in the auxiliary
two-dimensional theory in \cite{nair} and in terms of topological string
on twistor target space in \cite{witten}.
There are no amplitudes with zero or one negative
helicity gluons at the tree level however these amplitudes
emerge at one-loop level in the YM theory without supersymmetry
\cite{mahlon}. For instance, one-loop all-plus amplitude reads as
\beq
A^{one-loop}(+,\dots,+)= g^n \sum_{1\leq i_1<i_2<i_3<i_4\leq n}
\frac{\langle i_1,i_2\rangle[i_2,i_3]\langle i_3,i_4\rangle[i_4,i_1]}
{\langle1,2\rangle\langle2,3\rangle\dots \langle n,1\rangle}
\eeq

It was suggested in \cite{csw} that conventional YM diagrams in
both supersymmetric and non-supersymmetric gauge theories can be
reorganized in the different way which  nowadays is known as MHV
diagrams or CSW Lagrangian. The building blocks of this
diagrammatics are MHV vertices extended off-shell and the
canonical propagator $\frac{1}{P^2}$ involving $(+-)$ degrees of
freedom and connecting two MHV vertices. The continuation
off-shell suggested in \cite{csw} for $\lambda$ in any internal
line reads as \beq \lambda_{a}= p_{a,\dot{a}}\eta^{\dot{a}} \eeq
where $\eta$ is arbitrary spinor fixed for  off-shell lines in all
diagrams relevant for a given amplitude.

At higher loops the situation turns out to be more subtle at least
in the theory without supersymmetry. The non-vanishing all-plus
one-loop amplitude can not be derived from MHV vertices only that
is why it was suggested in \cite{one-loop} that one-loop  all-plus
diagram has to be added to the CSW Lagrangian as a new vertex. It
was also argued that there is no need to add one-loop vertex
$(-,+\dots +)$ to  new Lagrangian. The situation in SUSY case is
more safe since these amplitudes vanish however even in this case
it is unclear if new vertices have to be added to reproduce higher
loops results.

In spite of the considerable success  of this
approach its conceptual origin
remained obscure and it was unclear how these effective
degrees of freedom involved into the CSW Lagrangian are
related with the conventional YM gauge fields. It is
the goal of this paper to argue that these effective
degrees of freedom emerge from the standard YM variables
in the light-cone gauge upon the particular "dressing " procedure.

\section{Yang-Mills on the light cone}

In this Section we briefly discuss YM theory in the light-cone
gauge which involves only two physical degrees of freedom. The
Lagrangian of YM theory in the light-cone variables
has been found in N=4 SUSY case
\cite{mandelstam,brink}. In what
follows we shall exploit  Mandelstam two-field formulation \cite{mandelstam} which
has been  successfully used recently in one-loop calculations
in YM theories with different amount of supersymmetry \cite{bdkm}.
Two fields $\Phi_{+}$ and $\Phi_{-}$ are related with the
physical transverse degrees of freedom of the gluon as follows
\beq
\Phi_{-}(x)=\partial_{+}^{-1}A(x),\quad \Phi_{+}(x)=\partial_{+}{\bar{A}(x)}
\eeq

We shall be interested in the non-supersymmetric
theory with the action in $A_{+}=0$ gauge
$$
S= \int d^4x[\Phi_{+}^a\Box\Phi_{-}^a +2gf^{abc}
\partial_{+}\Phi_{-}^a\bar{\partial}\Phi_{-}^b\Phi_{+}^c +
2g f^{abc}\partial_{+}^{2}\Phi_{-}^a \partial_{+}^{-2}\partial\Phi_{+}^b
\partial_{+}^{-1}\Phi_{+}^c
$$
\beq
\label{n=0}
-2g^2f^{abc}f^{ade}
\partial_{+}^{-2}(\partial_{+}\Phi_{-}^{b}\Phi_{+}^{c})(\partial_{+}^{-1}\Phi_{+}^{d}
\partial_{+}^{2}\Phi_{-}^{e})]
\eeq
where $\partial=\frac{1}{\sqrt{2}}(\partial_{x_1} +i
\partial_{x_2})$ is derivative with respect to the transverse
coordinates $x_1,x_2$ and $\bar{\partial}=\partial^{*}$. The
action contains local and non-local triple vertices as well as
non-local quartic vertex.

Let us make a few comments on the form of the action (\ref{n=0}). First note
that it involves two fields of dimensions 0 and 2 hence positive
and negative helicity fields
enter Lagrangian asymmetrically. In particular, vertex $(-++)$ is local
in the coordinate space while $(--+)$ is not. There are two classes
of solutions to the equations of motion which correspond to the
self-duality and anti-self-duality equations written in  a little bit unusual
form, namely
\beq
\Phi_{-}=0 \qquad \Box \Phi_{+}= (\Phi_{+},\Phi_{+})
\eeq
and
\beq
\Phi_{+}=0 \qquad \Box \Phi_{-}= \{\Phi_{-},\Phi_{-}\}
\eeq
where the schematically written r.h.s. are obtained  by the variations
of the cubic terms in the action (\ref{n=0}).

Note  that
the truncation of the light-cone action to the
first two terms which
amounts to the self-dual equation of motion for
the negative helicity field has been discussed
in the context of MHV amplitudes in \cite{chalmers}.
However this truncation evidently can not be equivalent
to the full YM theory we are dealing with. The action (\ref{n=0})
is Gaussian with respect to both fields that is one of them
can be integrated out yielding highly nontrivial
effective action with a non-canonical kinetic term
for the other.

\section{MHV vertex from self-duality equation}

The essential ingredient of the MHV diagrams is
MHV vertex and in this section we shall argue that
solution to the self-duality equation
with the particular boundary conditions serves as
the generating function for all MHV amplitudes.
This fact has been recognized some time ago
by Bardeen \cite{bardeen} and has been elaborated
further in \cite{rs,cangemi,chalmers,korepin}. In what
follows just this solution to the self-duality equation
provides the desired change of variables from
conventional YM to MHV formalisms.

Let us briefly remind the derivation of the perturbiner
solution to the self-duality equation following \cite{rs}.
The self-dual perturbiner yields the form-factor of the one off-shell
gluon between the vacuum and arbitrary number of  gluons
of the same helicity, momenta $p_j$ and color orientations $t_j$.
The starting point
is the transition to the twistor representation with
the additional spinor homogeneous coordinate $\rho^{\alpha}$ on the
auxiliary $CP^1$. The self-duality equation in the twistor representation
is equivalent to the zero-curvature condition
\beq
[\nabla_{\dot{\alpha}}\nabla_{\dot{\beta}}]=0
\eeq
where $\nabla_{\dot{\alpha}}=\rho^{\alpha}\nabla_{\dot{\alpha},\alpha}$.
Hence the solution to the self-duality equation can be represented
in the following form
\beq
\label{ag}
A_{\dot{\alpha}}=g^{-1}\partial_{\dot{\alpha}}g
\eeq
where $\partial_{\dot{\alpha}}=\rho^{\alpha}\partial_{\dot{\alpha},\alpha}$ ,
$A_{\dot{\alpha}}=\rho^{\alpha}A_{\dot{\alpha},\alpha}$
and $g$ is group valued function depending on $\rho$ and $x$, as well as on
the quantum numbers $p_j$ and $t_j$ of the external particles.
We assume that
$A_{\dot{\alpha}}$ is a polynomial of degree one in $\rho$. Then the
group element necessarily has to be meromorphic function of $\rho$
of degree zero such that connection $A_{\dot{\alpha}}$ is regular
at the poles.

The perturbiner is defined as a solution to the self-duality equation
of the shape of a formal expansion in the
(non-commuting) variables $E_{j}=t_j e^{ip_{j}x}$, which are essentially the
plane waves of the external gluons of the same, say positive, helicity.
That is we look for the group element
providing the solution to the zero curvature equation in the
following form
\beq
g_{ptb}(\rho)=1 +\sum_{j}g_{j}(\rho)E_{j} +\dots +\sum_{j_1\dots j_{L}}
g_{j_1\dots j_{L}}(\rho)E_{j_1}\dots E_{j_{L}}+\dots ~,
\eeq
where different terms with $L$ of $E$'s correspond to different color orderings in the
form-factors with $L$ external particles.
The regularity of the connection at the poles leads us immediately
to a unique solution for coefficients of the
expansion of $g_{ptb}(\rho)$ \cite{rs}:
\beq
\label{pert}
g_{j_1\dots j_{L}}(\rho)=\frac{\langle\rho,q\rangle}
{\langle\rho,j_1\rangle\langle j_1,j_2\rangle\langle j_2,j_3\rangle\dots\langle j_{L-1},j_L\rangle}
\eeq
where the so-called reference spinor $q_\alpha$ is the one which enters
into the polarization vectors
$\epsilon_{\dot{\alpha},\alpha}^{j}=q_{\alpha}\tilde{\lambda}_{\dot{\alpha}}^{j}$.

The  corresponding connection $A_{ptb}$ can be found upon the substitution
of the solution into (\ref{ag}).
Let us note that perturbiner solution itself is localized on the
line in the twistor space if one performs  half-Fourier transforms
for all massless particles involved in the form-factor similar to
\cite{witten}.

The perturbiner solution describes form-factor or off-shell current of the form
$\langle A_{\dot\alpha,\alpha}(k)\rangle_{k_1,\dots,k_n}$
where the gluon with momentum $k$ is off-shell
while all other gluons are on-shell and have the same helicity. Using
the explicit form of the solution one can verify that this form-factor
has no pole in $k^2$ and, hence, gives zero upon the application of the reduction
formula. This corresponds to the vanishing of the amplitude with
all but one gluons of the same helicity.

To get  MHV amplitudes from the perturbiner solution one has to consider
the linearized YM equation in the background of the perturbiner. The most
compact form of the generating function for  MHV amplitudes  has the following
structure \cite{rs}
\beq
\label{action}
M(k_1,k_2)= \langle1,2\rangle^2\int d^{4}x Tr[E_{1} g_{ptb}^{-1}E_{2}g_{ptb}],
\eeq
where $k_1,k_2$ are momenta of the negative helicity gluons
and plane waves corresponding to the positive helicity
gluons are substituted into $g_{ptb}$. The group
elements which depend on the twistor variable have to be taken
at points  $\rho_i$ corresponding to the momenta of negative helicity gluons.

\section{Change of variables}

Let us turn to the central point of our paper and describe
the proper non-local change of variables. First, let us comment on the choice
we shall make in a moment. In the light-cone action
there is nontrivial $(-++)$ vertex which has to be absent
in the CSW Lagrangian. That is   change
of variables has to provide the removal of this term.
It enters the equation of motion for $\Phi_{+}$
which reduces to the self-duality equation if $\Phi_{-}=0$.
Hence we expect that change of variables we are
looking for should map self-duality equation
to the Laplace equation. Actually perturbiner
solution to the self-duality equation does this job.

To describe new variables precisely let us represent combination involved in
the equation of motion for the light-cone variable $\Phi_{+}$
at $\Phi_{-}=0$ in the  form
\beq
\Box\Phi_{+} +(\Phi_{+},\Phi_{+})=\partial_{+}F'(\phi_{+},\partial_{+}^{-1}\Box \phi_{+})
\eeq
where the following notation is assumed  $\delta F=F'(\phi,\delta\phi)$.
Now introduce new variable $\phi_{+}$ by
\beq
\Phi_{+}=F(\phi_{+})
\eeq
The convenient choice for the  second field $\phi_{-}$ is dictated
by the canonicity of the $(+-)$ propagator in new
variables which yields
\beq
\Phi_{-}=\partial_{+}^{-4}F'(\phi_{+},\partial_{+}^4\phi_{-})
\eeq
The correct form of the propagator can be checked with the
help of relation
\beq
\int d^{4}x Tr[F'(\phi,v)\partial_{+}^{-3}F'(\phi,u)]= \int d^{4}x Tr[u\partial_{+}^{-3}v]
\eeq
valid for arbitrary $u$ and $v$.

Now we are ready to make a link with the previous Section. It
turns out that \beq \label{change} F(\phi_{+})=\partial_{+}
g_{ptb}^{-1}(\phi_{+})\partial g_{ptb}(\phi_{+}) \eeq where  group
element in (\ref{change}) is effectively continued off-shell. That
is the off-shell field $\phi_{+}$ is considered  in $g_{ptb}$
instead of the plane wave  and momenta of plane waves are
substituted by the corresponding derivatives
\beq
\Phi_{+}=\partial_{+}^2 \sum_{n\geq 1}
\frac{1}{{\partial}_{+,1}\langle\bar{\partial_{1}}\bar{\partial}_{2}\rangle
\dots {\partial}_{+,n}}\underbrace{\phi_{+}...\phi_{+}}_{n}
\eeq
where $\bar{\partial}_{k}=\partial_{\alpha,k}$ acts on the k-th
term in the product. Effectively the change of variables above
kills $(++-)$ vertex in the action and maps solution to the
self-duality equation to the solution to the free Laplace equation.
The rest of the check concerns the interaction part of the
Lagrangian. We have verified that $(--+)$ and $(++--)$ vertices in
the light-cone Lagrangian get combined together into the correct
interaction terms in CSW Lagrangian. That is we have argued that
change of variables from light-cone YM fields yields at the tree
level both correct propagator and vertices in the CSW Lagrangian.
Hence just fields $\phi_{-}\phi_{+}$ play the role of twistor
degrees of freedom corresponding to positive and negative
helicities. The technical details concerning the change of
variables shall be presented elsewhere.

Let us make a few comments on the one-loop extension of the CSW Lagrangian.
As we have already mentioned in
the non-supersymmetric case it has to be extended
by all-plus one-loop amplitude. The possible origin of such correction is
clear in our approach - there is Jacobian of the change of variables.
We have not proved  that  Jacobian reproduces the desired
answer but there are several arguments favoring this possibility.
Naively the change of variables discussed is expected to be canonical that
is if we would work with the system with finite number degrees of freedom
then it would equal to one. However the theory at hands enjoys
infinite dimensional phase space hence one could
expect the anomalous Jacobian of the canonical transformations.
Second argument concerns the form of the naive Jacobian which
involves  only powers of $\phi_{+}$ as expected.
Moreover if the additional one-loop term follows from the Jacobian indeed then
the absence of such terms in SUSY case could be attributed naturally
to the standard SUSY cancellations.

Note that in principle the second similar change of variables
can be done which would kill the $(--+)$ vertex as well. Upon
this change the action would involve only quartic and higher
interaction terms however the possible usefulness of such
action is unclear to us at present.

\section{Discussion}

In this short note we questioned the relation
between the conventional YM variables and
effective degrees of freedom in the CSW
Lagrangian. The answer turns out to be remarkably
simple - they are related just by the non-local
change of variables. Tree diagrams are perfectly
reproduced upon this change while the evident
candidate for the one-loop completion of the action
is the corresponding Jacobian. Moreover our
consideration implies that one should not expect additional
terms in the CSW Lagrangian at higher loops.

The immediate question in our approach
concerns the twistor interpretation of the
suggested change of variables. We expect that the interpretation
of the perturbative YM theory in terms of the string
on the twistor manifold \cite{witten}
matches the twistor interpretation of the
perturbiner developed in \cite{rs}. Of course the most
interesting question raised in \cite{bardeen} concerning
the expected relation with some hidden integrability-like
structure responsible for the nullification of the
infinite number of tree amplitudes
remains open. Nevertheless we believe that our work
could be useful for this line of reasoning.

There are several possible generalizations.
First, supersymmetric case can be considered along
this way starting with the corresponding light-cone
formulation \cite{mandelstam}. The perturbiner solution
in the supersymmetric case has been found in \cite{selivanov}.
It is known as well for the gravity case \cite{rs2}.
One more line of generalization concerns QED with
massless or massive fermions.  Recently
MHV-type technique for tree QED was developed
\cite{stirling} which
naturally captures the soft photons limit while the
example of the stringy picture for MHV QED amplitude
has been found in \cite{gl}. The
change of variables could be found in this case as well
and we expect that effective fermionic fields in MHV formulation
of QED involve original fermions dressed by the infinite
number of positive or negative helicity photons.

We are grateful to A. Gerasimov, V.A. Khoze and
A.Vainshtein for the useful discussions. The work
of A.G. was supported in part by grants CRDF RUP2-261-MO-04
and RFBR-04-011-00646 and work of A.R. by grants RFBR-03-02-17554,
NSch-1999.2003.2 and  INTAS-03-51-6346.
A.G. thanks FITP Institute at University of Minnesota
where the paper has been completed for the kind hospitality and support.

\end{document}